\documentclass[pre,twocolumn,showpacs,preprintnumbers,amsmath,amssymb,floatfix]{revtex4}
\def\beneq{\begin{equation}}
\def\eneq{\end{equation}}
\def\bea{\begin{eqnarray}}
\def\eea{\end{eqnarray}}
\usepackage{graphicx}
\usepackage{dcolumn}
\usepackage{bm}
\begin{document}

\title{Combined macro- and micro-rheometer for use with Langmuir monolayers}

\author{Robert Walder,$^{1}$, Christoph
Schmidt$^{2}$, and Michael Dennin$^{1}$}

\affiliation{
   $^{1}$Department of Physics \& Astronomy,
University of California, Irvine, CA 92697\\
$^{2}$Drittes Physikalisches Institut, Fakult\"{a}t f\"{u}r
PhysikGeorg-August-Universit\"{a}t, 37077 G\"{o}ttingen, Germany }

\date{\today}

\begin{abstract}
A Langmuir monolayer trough that is equipped for simultaneous
microrheology and standard rheology measurements has been
constructed. The central elements are the trough itself with a
full range of optical tools accessing the air-water interface from
below the trough and a portable knife-edge torsion pendulum that
can access the interface from above. The ability to simultaneously
measure the mechanical response of Langmuir monolayers on very
different length scales is an important step for our understanding
of the mechanical response of two-dimensional viscoelastic
networks.
\end{abstract}


\maketitle

Soft-matter systems play critical roles both in technological
applications and in biology. Most examples of soft matter are
complex fluids, exhibiting both solid- and fluid-like responses.
One of the challenges in soft-matter research is the quantitative
understanding of such behavior. Complex dynamics are typically the
result of complex structure spanning a range of length scales.
Often, the large-scale response of the system depends in a
non-trivial fashion on the interaction between mesoscopic
structures in the system. The local response within one of these
mesoscopic structures can be very different from the macroscopic
response. For example, aqueous foam is composed of gas bubbles
separated by liquid films. Measurements within a liquid film
suggest the material is fluid. However, for applied stress below
the yield stress, the macroscopic foam acts as a solid. Due to
various additives, the liquid film itself may be viscoelastic, and
the details of this ``microscopic" response can, in turn, directly
impact the details of the macroscopic behavior.

A system of great biological interest, for which length scales
play an important role are semi-flexible polymer networks. Many
structures in biology are composed of semi-flexible networks of
protein filaments, such as the actin cytoskeleton, that are also
connected with lipid membranes. The need to understand both the
microscopic and macroscopic response of such a composite system
provided much of the motivation for the apparatus we report on in
this paper. The other motivation was the need to make measurements
on a robust model system. For this, we have chosen to focus on
Langmuir monolayers as a model for the lipid structures. Langmuir
monolayers are single-molecule-thick layers of amphiphilic
molecules at the air-water interface, and thereby resemble a half
of a lipid bilayer which forms cell membranes. There exists a
well-developed set of tools for studying their mechanical
properties. This offers many advantages when pursuing mechanical
studies, and it allows us to build on existing technology.

One can categorize the existing methods for measuring the
viscoelastic properties of Langmuir monolayers according to the
length scale probed by the technique. The more common techniques
that probe the system on relatively large length scales include
needle viscometers \cite{BFFR99}, channel viscometers
\cite{HK38,CHB42,SYZ93,SKB94}, the knife-edge torsion pendulum
\cite{MS70,FMA92,AMXK83,KSMBS94,GKMD97}, and light-scattering
\cite{HL76,L81}. Our group has also developed a Couette-style
rheometer that expands on standard knife-edge torsion pendulum
techniques \cite{GD98}.

The first use of microscopic-scale local techniques in Langmuir
monolayers was based on measuring relations between translational
and rotational mobilities of particles diffusing in the monolayer
\cite{HPWS82,SD75}. In this regard, a useful feature of Langmuir
monolayers is their rich phase behavior, with the equivalent of
gas, liquid, liquid crystal (often referred to as liquid
condensed), and crystalline phases. Because they often exhibit
phase coexistence, the domains of one phase can be used as the
`particles'' to probe the properties of the other phase. For
example, experiments have included measuring the diffusion of
liquid crystalline domains \cite{KM93}, and using liquid
crystalline domains as particles for measuring velocity profiles
\cite{SKB94}. More recently, Brewster-angle microscopy has been
used to study the motion of the domains \cite{S01,IIS01}. Our
apparatus uses microrheological techniques that rely on the
introduction of tracer particles. It is based on similar
techniques discussed in Ref.
\cite{Fischer:00,Fischer:01a,Fischer:01b,Rondelez:03}. However, as
discussed in more detail later, we expand on these techniques in
several ways.

\begin{figure*}
\centering
\includegraphics[width=400pt]{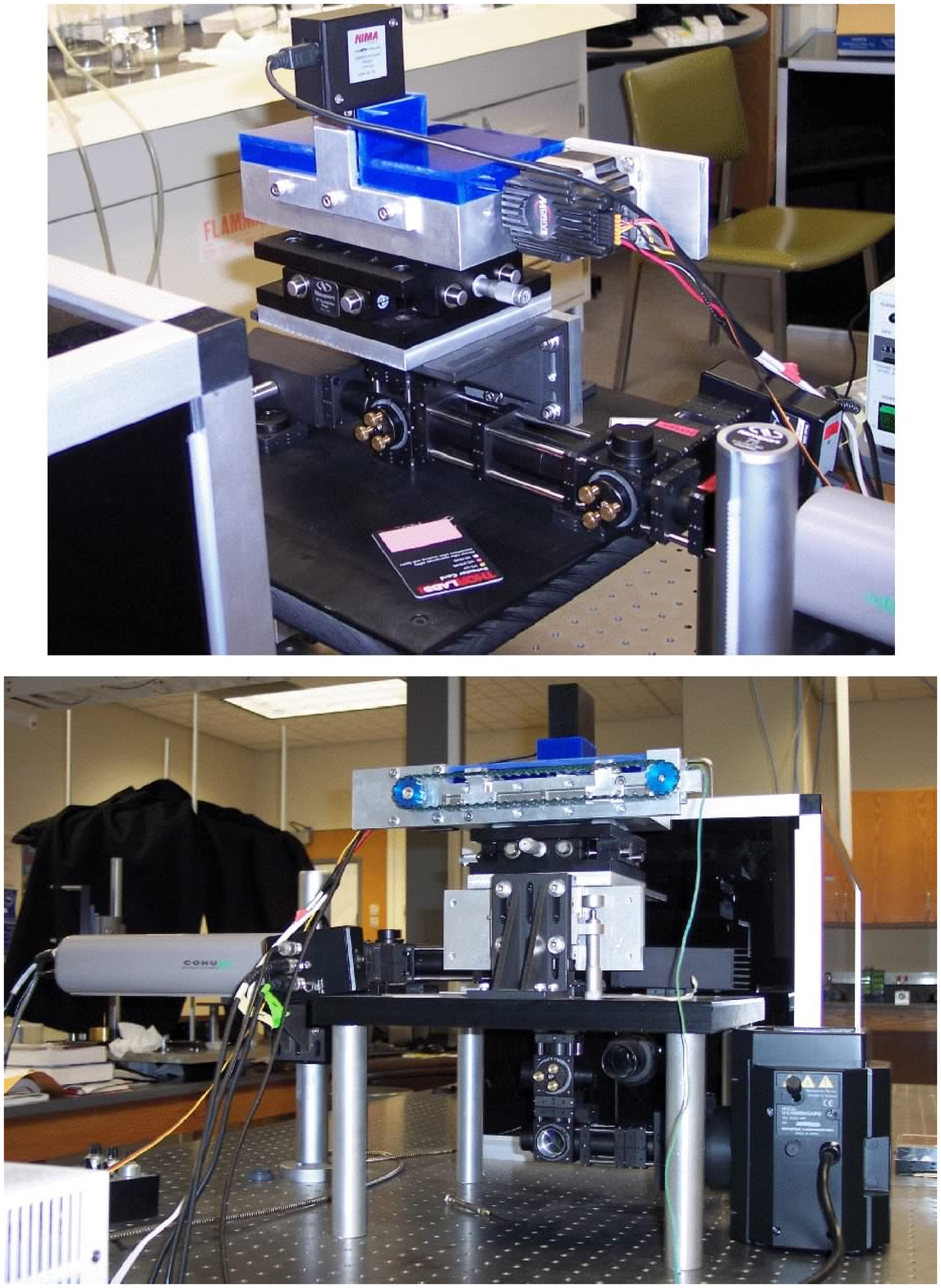}
\caption{(color online) Two pictures of the apparatus. The top
view highlights the view from where the laser enters the optics
below the trough. The laser comes from the lower left portion of
the image. The lower image is the view from the opposite side of
the trough. The black case in the background houses the laser.
(For a scale, the spacing between the holes on the optical table
are one inch.)}\label{photo}
\end{figure*}

\begin{figure*}
\centering
\includegraphics[width=400pt]{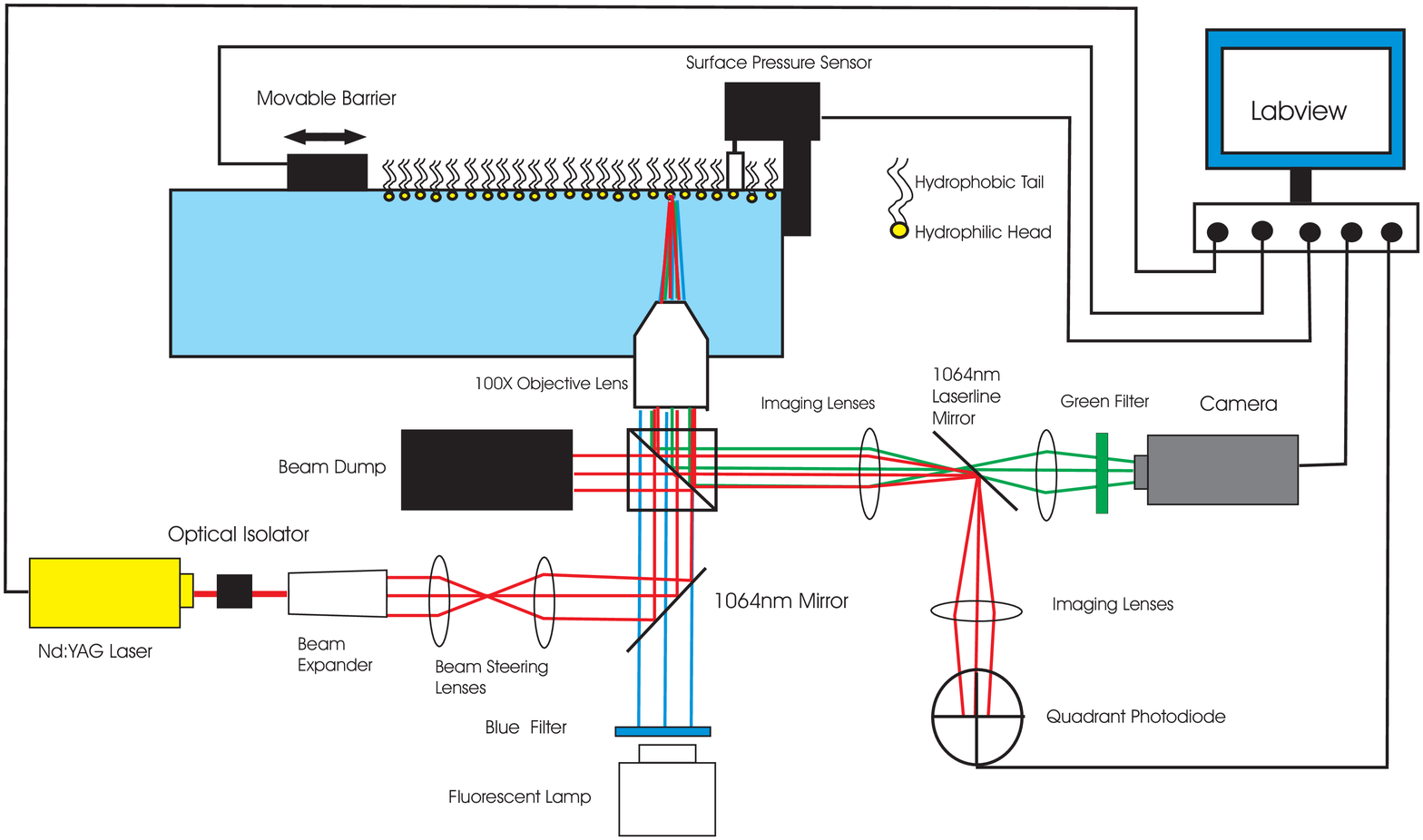}
\caption{(color online) Entire Experimental setup}\label{Full
Experiment}
\end{figure*}

In considering the design of an apparatus that can measure
rheological properties on multiple length scales, we decided to
pursue a  modular design that combines two established rheological
techniques: particle microrheology and the knife-edge torsion
pendulum. Both of these measure the complex shear modulus, $G^*$.
In linear response, the complex shear modulus provides the
connection between an oscillatory strain $\gamma(\omega)$ and
stress $\sigma(\omega)$ as a function of the oscillation frequency
$\omega$ through the relation: $\sigma(\omega) = G^*
\gamma(\omega)$. The real part of $G^*$ is the elastic modulus and
the imaginary part is related to the viscous response (i.e. energy
dissipation in the system). The difference between the
microrheology measurements and the knife-edge torsion pendulum is
in the length scales being probed. Typically, microrheology uses
micron-size beads embedded in the material and the torsion
pendulum is on the order of mm to cm. We will briefly review each
technique and discuss the design goals for each of these. Then, we
will describe our apparatus in detail.

The most basic form of microrheology uses position fluctuations of
tracer particles (typically micron sized spheres) embedded in the
material as a probe \cite{Schmidt:97,SchmidtPRL:97,Schmidt:99}.
The displacement of a single spherical probe is described by the
standard Stokes-Einstein relation\cite{SchmidtPRL:97,Levine:00}:
\[u(\omega) = \frac{f(\omega)}{6\pi a G(\omega)}\]
where $G(\omega) = - i\omega\eta$ is the complex shear modulus.

This method is termed passive microrheology because only the
thermal force fluctuations experienced by the particles are used
as the driving force. Our system is based on a procedure for
determining the shear moduli that is fully discussed in Schnurr et
al \cite{Schmidt:97}. Essentially, one uses the
fluctuation-dissipation theorem to relate the power spectral
density of the thermal motion of the bead to the imaginary part of
the complex response function $\alpha''(f)$. When measuring
$\alpha''(f)$ over a large enough frequency space, a
Kramers-Kronig integral can be used to recover the real part of
the response function $\alpha'(f)$, and the total complex response
function $\alpha(f)$ is then related to the complex shear modulus
$G(f) = G'(f) + \imath G''(f)$ by the generalized Stokes-Einstein
relation (GSER):

\[G(f) = \frac{1}{6 \pi a \alpha(f)}\].

Therefore, a critical design element in the system is the ability
to measure over a wide range of frequencies. This will be
discussed in the detailed description of the apparatus. The final
key experimental component for the microrheology is the use of
optical tweezers to trap the probe particle \cite{Chu:86}.

The knife-edge torsion pendulum method is well established to
probe viscoelastic properties of surface layers on air-water
interfaces. Essentially, one suspends a Teflon disk with a
knife-edge at its circumference in the monolayer. The disk is
driven with a known applied torque, which corresponds to an
applied stress for the monolayer. The angular response of the disk
corresponds to the strain response of the monolayer. One must
correctly account for the contribution of the fluid-subphase, and
the details of this are given in Ref.~\cite{GKMD97}. The key
elements in this case are the driving mechanism and the
measurement of the rotation angle. For the application discussed
here, it was necessary to adopt a design that could be easily
integrated with the microrheology technique.

To achieve the integration of the knife-edge torsion pendulum with
the microrheology technique required the design of a new Langmuir
monolayer trough with full optical access from below and the
adaptation of our existing knife-edge torsion pendulum. Access
from below was necessary for the optical tweezers and particle
tracking in order to harness the full power of microrheology. In
addition, for added flexibility, we provided access from below for
a fluorescence microscope so as to leave space above the monolayer
for the knife-edge torsion pendulum. The main mechanical
adaptation necessary for the knife-edge torsion pendulum was the
design of a modular unit that can function as a stand-alone
device. This makes it possible to remove the device for the use of
additional tools in combination with the optical tweezers. It has
the added benefit that the knife-edge torsion pendulum can be used
and calibrated in a separate mini-Couette rheometer. We will first
describe the Langmuir trough and integrated microrheology
apparatus and then we will separately discuss the design of the
stand-alone knife-edge torsion pendulum. The combination of the
two techniques into one instrument creates a powerful,
high-precision instrument capable of exploring mechanical
properties of surface films that were previously not accessible.

The integrated Langmuir trough consists of three optical
instruments and a surface tensiometer. Figure~\ref{photo} provides
two view of the apparatus, and Fig.~\ref{Full Experiment} is a
schematic of the overall apparatus. Details of each portion of the
apparatus are provided in the following sections. The integration
of these systems into one master instrument allows us to
simultaneously image and mechanically manipulate surface films in
the Langmuir trough and track embedded particles. A number of the
components are standard equipment for Langmuir monolayer troughs.
For example, a commercial surface pressure sensor
\cite{SurfacePressureSensor} is used to characterize the phase of
the monolayer. A typical isotherm is illustrated in Fig.~\ref{DPPC
Isotherm} for a DPPC monolayer.

\begin{figure} [h]
\centering
\includegraphics[width=225pt]{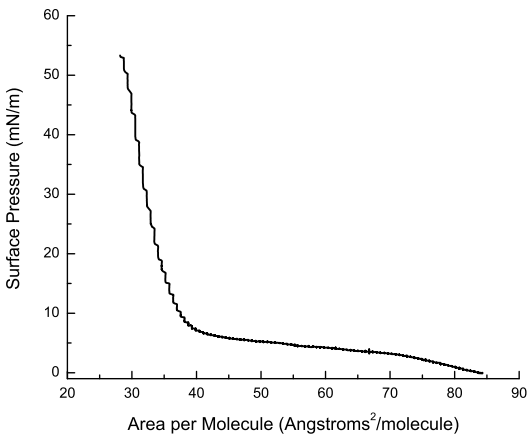}
\caption{Isotherm perfomed on DPPC monolayer at room
temperature}\label{DPPC Isotherm}
\end{figure}

\begin{figure} [h]
\centering
\includegraphics[width=225pt]{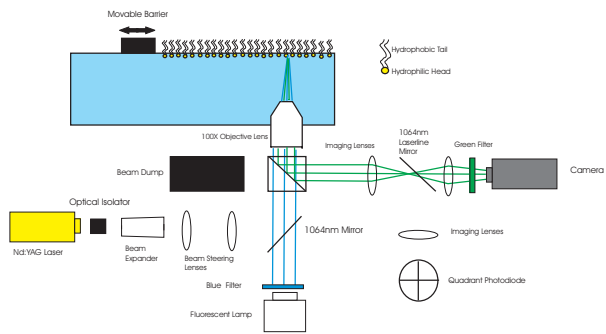}
\caption{(color online) Fluorescence
Microscope}\label{Fluorescence Microscope}
\end{figure}

Due to space and cost constraints, we have chosen to use a custom
built fluorescence microscope that images the monolayer from below
instead of mounting the system on a full commercial fluorescence
microscopic (see Fig.~\ref{Fluorescence Microscope}). Fluorescence
microscopy is essential to be able to specifically detect
molecules at the interface that can not be detected using standard
bright-field microscopy. Selected populations of molecules in the
sample (in our case lipids and/or actin) are modified by
chemically attaching a fluorophore.

In our optical configuration, we use a mercury arc lamp
\cite{Lamp} combined with appropriate excitation filters mounted
in a filter wheel, to excite fluorescence, and the corresponding
emission filters on a filter wheel connected to our intensified
CCD camera \cite{CCDCamera} to detect fluorescence. This
configuration allows us to use different fluorophores during our
experiments and thus allows us to selectively image different
components of our sample. The 2 filter cubes we have currently
installed are the FTIC and rhodamine filters.  One particularly
useful application of this capability is to use one fluorophore
for the lipid monolayer (such as Bodipy) and to use another
fluorophore for proteins introduced into the subphase (such as
rhodamine actin).

\begin{figure} [h]
\centering
\includegraphics[width=225pt]{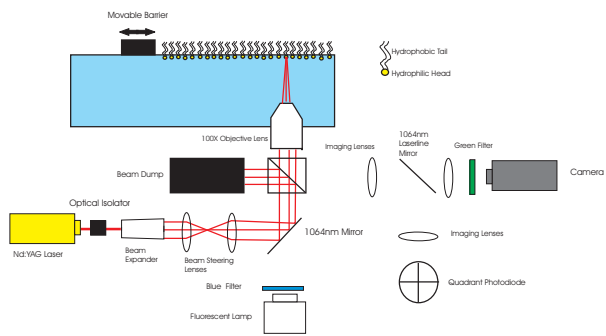}
\caption{(color online) Optical Tweezer}\label{Optical Tweezer}
\end{figure}

The basic principle of an optical tweezer is to focus a reasonably
high-powered laser to a diffraction limited focus. The light
intensity gradient creates a force on objects of a higher index of
refraction than the surroundings. The two basic components of an
optical tweezers are thus a laser \cite{Laser} and an objective
lens \cite{ObjectiveLens} (see Fig.~\ref{Optical Tweezer}).  The
objective lens should have a high numerical aperture to create a
steep enough intensity gradient \cite{LANG03}.  The water
immersion objective lens we use for our optical tweezer system has
a numerical aperture of 1.0 and a working distance of 1.5~mm.

\begin{figure} [h]
\centering
\includegraphics[width=225pt]{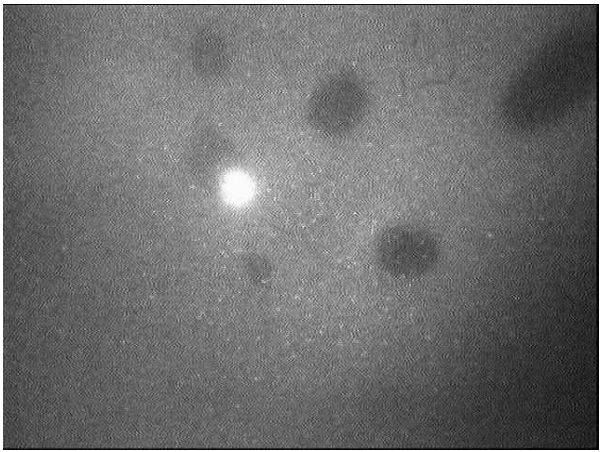}
\caption{Imaging a trapped bead in a DPPC monolayer at a surface
pressure of 6mN/m.}\label{DPPC Image}
\end{figure}

Additional optical components were added to the laser/objective lens
system in order to control trap strength and to allow us to steer
the trap location in the sample.  An optical isolator prevents
backscattering which can cause laser instability.  A beam expander
increases the beam diameter 4-fold in order to use the maximal
numerical aperture by filling the back aperture of the objective
lens. Two 100~mm focal length lenses form a telescope for beam
steering. The first lens is mounted to an 3-axis translation stage.
Translation of this lens steers the trap location in 3 dimensions
within a range of several microns in each axis. An example of an
image of a particle in the trap (using the Fluorescence microscope)
is shown in Fig.~\ref{DPPC Image}

\begin{figure} [h]
\centering
\includegraphics[width=225pt]{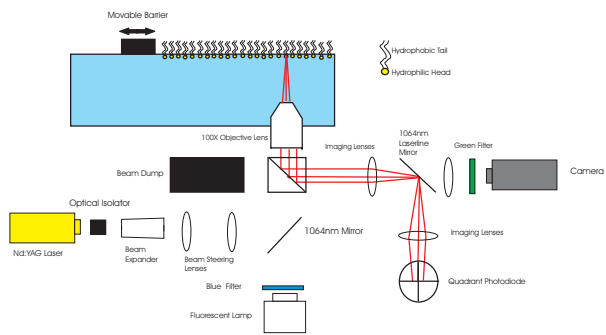}
\caption{(color online) Particle Tracking}\label{Particle
Tracking}
\end{figure}

A key element of the microrheology system is the particle
displacement measurement. The thermal position fluctuations are
close to and beyond the limits of the spatial optical resolution
that can be achieved with standard video particle tracking. In
addition, the frame rate of the CCD camera used for fluorescence
microscopy is limited to a range of 10-100~Hz. This imposes a limit
on the range of frequencies over which the complex shear modulus can
be determined by video particle tracking. Using the quadrant diode
based displacement detection system described in the next section
(see Fig.~\ref{Particle Tracking}), we can probe the frequency
response of the system up to 100~kHz.

The back-scattered laser light from the trapped particle is imaged
onto a quadrant photodiode in a method similar to Helfer et al
\cite{Chatenay:00}. When forward-scattered light can be collected,
one can use back-focal plane interferometry
\cite{Allersma:98,Schmidt98}. In our case, we have no optical
access from above and therefore use backscattered light. The
detection optics consists of a beam splitter, imaging lenses, a
$45^0$ degree 1064~nm laser line mirror, and a quadrant photodiode
\cite{QuadrantPhotodiode}. The beam splitter is used to reflect
some of the back-scattered laser light from the trapped particles
into the imaging optics.  The lenses are positioned to form an
image of the particle on the quadrant photodiode with a final
magnification of 100X. The laser line mirror is used both as a
filter to prevent any visible light from the fluorescence imaging
from reaching the quadrant photodiode and to center the
back-scattered light onto the quadrant photodiode.  In addition,
the laser line mirror protects the CCD camera from being damaged
by the backscattered laser light. The quadrant photodiode then
records the position of the particle with high spatial and
temporal resolution in terms of the four output photo-currents.
The sampling frequency is typically 66~kHz and the number of
points sampled is $2^n$, in order to use fast fourier methods for
data analysis.

The use of backscattered laser light means that we are photon
limited and that our measurements can not quite achieve the
performance of interferometric based
systems\cite{Schmidt:98,Tang:04}. However, this optical system
allows us to expand on previous measurements of static viscosity
and drag coefficient measurements for monolayers
\cite{Fischer:01a,Rondelez:03} into a significantly larger
frequency range. A set of data is shown in Figs.~\ref{Power
Spectral Density},\ref{G Prime}, and \ref{G2 Prime} for the power
spectral density, elastic modulus, and loss modulus, respectively
of a one micron diameter bead in a DPPC monolayer at a surface
pressure of 6~mN/m.

\begin{figure} [h]
\centering
\includegraphics[width=225pt]{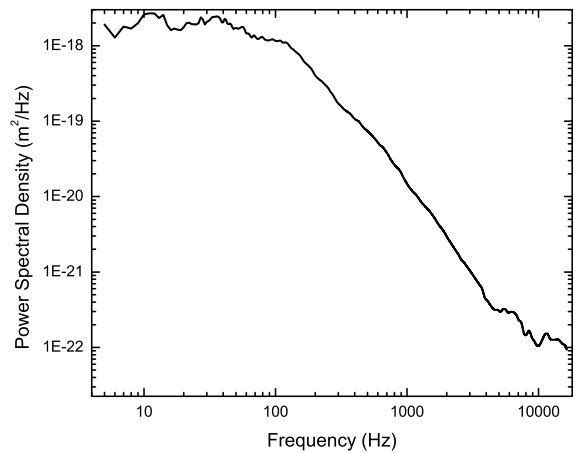}
\caption{Power Spectral Density of a DPPC monolayer at a surface
pressure of 6 mN/m. The data has been smoothed by logarithmic
averaging.}\label{Power Spectral Density}
\end{figure}

\begin{figure} [h]
\centering
\includegraphics[width=225pt]{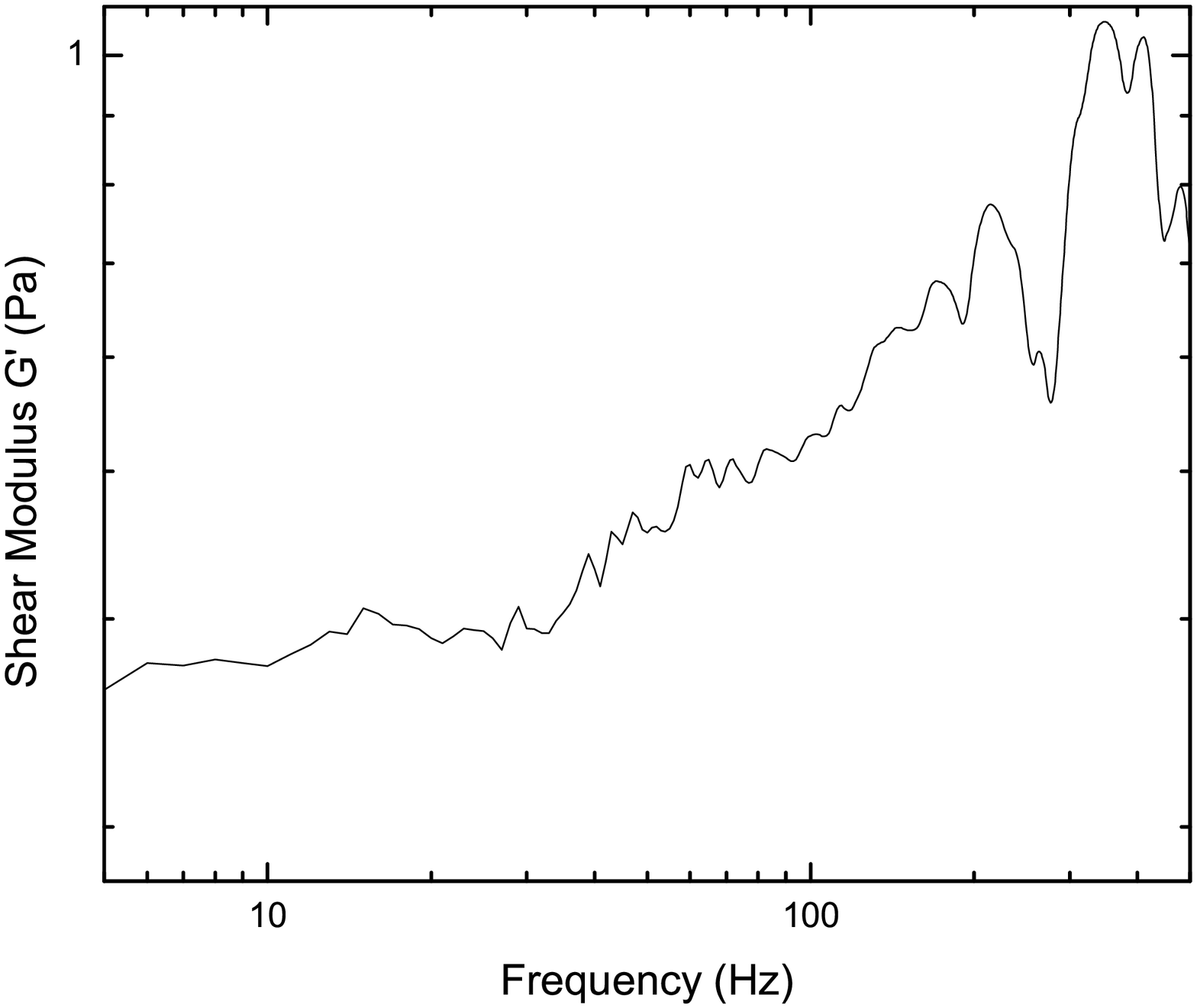}
\caption{Elastic Modulus of a DPPC monolayer at a surface pressure
of 6 mN/m. This has been calculated from the power spectral
density shown in Fig.~\ref{Power Spectral Density}.}\label{G
Prime}
\end{figure}

\begin{figure} [h]
\centering
\includegraphics[width=225pt]{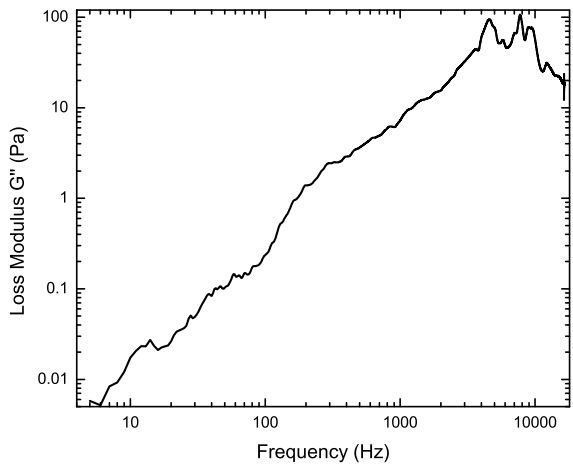}
\caption{Loss Modulus of a DPPC monolayer at a surface pressure of
6 mN/m. This has been calculated from the power spectral density
shown in Fig.~\ref{Power Spectral Density}.}\label{G2 Prime}
\end{figure}

\begin{figure}
\includegraphics[width=225pt]{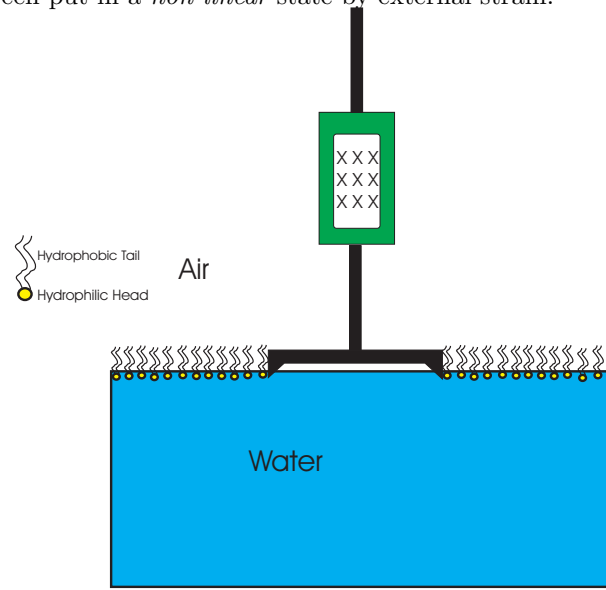}
\caption{(color online) Schematic of the portable torsion
pendulum.} \label{torsion}
\end{figure}

Because we have designed the main elements of the microrheological
methods to probe the system from below the trough, we can access
the surface of the water with a standard torsion, knife-edge
device. For this, we have constructed a removable device.
Figure~\ref{torsion} shows a schematic. The key elements are the
suspension of a circular Teflon disk by a torsion wire. The
support for the disk also contains a magnet coil (green rectangle
in Fig.~\ref{torsion}). This coil is situated within a larger coil
that generates a magnetic field, indicated by the x's in
Fig.~\ref{torsion}. There is additionally a fixed magnet attached
to the coil on the disk support.

The coils serve a dual purpose. With the coils, we are able to
both measure the angular position of the disk and provide an
external torque to the disk to drive it at a defined frequency.
The details of measuring the complex shear modulus with this
method are provided in Ref.~\cite{GKMD97}. The coils and torsion
pendulum match the specifications reported in Ref.~\cite{GD98}. We
have modified the support structure to allow for portability of
the device.

The primary motivation for building this apparatus was the need to
measure properties of Langmuir monolayers and monolayer-protein
complexes simultaneously on multiple length scales. It is
worthwhile to briefly discuss another important capability of this
system. Using the Teflon disk, one can apply a known macroscopic
strain to the system. One can then use the microrheology
measurements to determine the effect of strain on the local linear
response of the samples. In other words, one can probe the {\it
linear} response of a system that has been put in a {\it
non-linear} state by external strain.

\section{Acknowledgements}
We acknowledge the support of NSF-DMR-0354113. We also thank
Thomas Fischer for helpful discussions. R. Walder acknowledges
support through a travel fellowship from the Institute for Complex
Adaptive Matter. C.F. Schmidt acknowledges support by the DFG/
SFB755 Nanoscale Photonic Imaging and by the DFG Center for
Molecular Physiology of the Brain (CMPB).


\begin{thebibliography}{99}

\bibitem{BFFR99}
C.\ F.\ Brooks, G.\ G.\ Fuller, C.\ W.\ Frank and C.\ R.\ {\bf
15}, 2450 (1999).

\bibitem{HK38}
W.\ D.\ Harkins and J.\ G.\ Kirkwood, J Chem Phys {\bf 6}, 53
(1938).

\bibitem{CHB42}
L.\ E.\ Copeland, W.\ D.\ Harkins and G.\ E.\ Boyd, J Chem Phys
{\bf 10}, 357 (1942).

\bibitem{SYZ93}
M.\ Sacchetti, H.\ Yu and G.\ Zografi, Rev Sci Instr {\bf 64},
1941 (1993).

\bibitem{SKB94}
D.\ K.\ Schwartz, C.\ M.\ Knobler and R.\ Bruinsma, Phys Rev Lett
{\bf 73}, 2841 (1994).

\bibitem{MS70}
R.\ J.\ Mannheimer and R.\ S.\ Schechter,  J. Coll. Inter. Sci.
{\bf 32}, 195 (1970).

\bibitem{FMA92}
S.\ S.\ Feng, R.\ C.\ MacDonald and B.\ M.\ Abraham, Langmuir {\bf
7}, 572 (1992).

\bibitem{AMXK83}
B.\ M.\ Abraham, K.\ Miyano, S.\ Q.\ Xu and J.\ B.\ Ketterson, Rev
Sci Instr {\bf 54}, 213 (1983).

\bibitem{KSMBS94}
J.\ Kr"gel, S.\ Siegel, R.\ Miller, M.\ Born and K.-H.\ Schano,
Colloids and Surfaces A {\bf 91}, 169 (1994).

\bibitem{GKMD97}
R.\ S.\ Ghaskadvi, J.\ B.\ Ketterson, R.\ C.\ MacDonald and P.\
Dutta, Rev Sci Instr {\bf 68}, 1792 (1997).

\bibitem{HL76}
S.\ H\.{a}rd and H.\ L\"{o}fgren, Journal Colloids and Interface
Science {\bf 60}, 529 (1976).

\bibitem{L81}
D.\ Langevin, Journal Colloid Interface Science {\bf 80}, 412
(1981).

\bibitem{GD98}
R.\ S.\ Ghaskadvi and M.\ Dennin, Review of Scientific Instruments
{\bf 69}, 3568 (1998).


\bibitem{HPWS82}
B.\ D.\ Hughes, B.\ A.\ Pailthorpe, L.\ R.\ White and W.\ H.\
Sawyer, Biophys. J. {\bf 37}, 673 (1982).

\bibitem{SD75}
P.\ G.\ Saffman and M.\ Delbruck, Proc. Nat. Acad. Sci. USA {\bf
72}, 3111 (1975).

\bibitem{KM93}
J.\ F.\ Klingler and H.\ M.\ McConnell, Journal of Physical
Chemistry {\bf 97} 6096 (1993).

\bibitem{S01}
P.\ Steffen, et al., Journal of Chemical Physics {\bf 115}, 994
(2001).

\bibitem{IIS01}
A.\ T.\ Ivanova, J.\ Ignes-Mullol, and D.\ K.\ Schwartz, Langmuir,
{\bf 17}, 3406 (2001).

\bibitem{Fischer:00}
S.\ Wurlitzer, P.\ Steffen, and Th.~M.\ Fischer, Journal Of
Chemical Physics {\bf 112}, 13 (2000).

\bibitem{Fischer:01a}
P.\ Steffen, P.\ Heinig, S.\ Wurlitzer, Z.\ Khattari, and Th.~M.\
Fischer, Journal Of Chemical Physics {\bf 115}, 2 (2001).

\bibitem{Fischer:01b}
S.\ Wurlitzer, C.\ Lautz, M.\ Liley, C.\ Duschl, and Th.~M.\
Fischer, J. Phys. Chem. B {\bf 105}, 182 (2001).

\bibitem{Rondelez:03}
M.\ Sickert, and F.\ Rondelez, Physical Review Letters {\bf 90},
12 (2003).

\bibitem{Schmidt:97}
B.\ Schnurr, F.\ Gittes, F.C.\ MacKintosh, and C.F.\ Schmidt
Macromolecules {\bf 30}, 7781 (1997).

\bibitem{Levine:00}
A.J.\ Levine and T.C.\ Lubensky, Physical Review Letters {\bf 85},
1774 (2000).

\bibitem{SchmidtPRL:97}
F.\ Gittens, B.\ Schnurr, P.~D.\ Olmstedt, F.~C.\ MacKintosh, and
C.~F.\ Schmidt, Phys. Rev. Lett {\bf 74}, 3286 (1997).

\bibitem{Schmidt:99}
F.~C.\ MacKintosh and C.~F.\ Schmidt, Curr. Opin. Collid and
Interf. Sci. {\bf 4}, 300 (1999).

\bibitem{Chu:86}
A.\ Ashkin, J. M.\ Dziedzic, J. E.\ Bjorkholm, and S.\ Chu, Optics
Letters {\bf 11}, 5 (1986).

\bibitem{SurfacePressureSensor}
\emph{Nima Technology Ltd} Surface Pressure Sensor Type PS4, The
Science Park, Coventry, CV4 7EZ, England

\bibitem{Lamp}
\emph{Olympus America Inc.}, Model 5-UL155,  3500 Corporate
Parkway P.O. Box 610 Center Valley, PA

\bibitem{CCDCamera}
\emph{Cohu} Intensified CCD Camera Mod. 5515-2001, 3912 Calle
Fortunada, San Diego, CA 92123-1827

\bibitem{Laser}
\emph{Spectra Physics} Model BL-106C, 1335 Terra Bella Avenue
Mountain View, CA 94039

\bibitem{ObjectiveLens}
\emph{Olympus America Inc.}, Model 1-UM575,  3500 Corporate
Parkway P.O. Box 610 Center Valley, PA

\bibitem{LANG03}
M.\ J.\ Lang, and S.\ M.\ Block, Amer. Jour. of Phys. {\bf 71},
201 (2003).

\bibitem{Chatenay:00}
E.\ Helfer, S.\ Harlepp, L.\ Bourdieu, J.\ Robert, F.C.\
MacKintosh, and D.\ Chatenay, Physical Review Letters {\bf 85}, 2
(2000).

\bibitem{QuadrantPhotodiode}
\emph{New Focus} Model 2309, 2584 Junction Avenue San Jose, CA
95134.

\bibitem{Allersma:98}
M.\ W.\ Allersma and F.\ Gittes and M.\ J.\ Castro and R.\ J.\
Stewart adn C.\ F.\ Schmidt, Biophys. Jour. {\bf 74}, 1074 (1998).

\bibitem{Schmidt:98}
F.\ Gittes and C.\ F.\ Schmidt, Optics Letters {\bf 23}, 7 (1998).

\bibitem{Tang:04}
K.\ Addas, C.\ F.\ Schmidt, and J.\ Tang, Physical Review E {\bf
70}, (2004).


\end{thebibliography}
\end{document}